\documentstyle[12pt]{article}

\begin{document}
\title{Test Matter in a Spacetime with Nonmetricity}

\author{Yuval Ne'eman \\
Raymond and Beverly Sackler Faculty of Exact Sciences \\
Tel-Aviv University, Tel-Aviv, Israel 69978 \\
and \\
Center for Particle Physics, University of Texas, \\
Austin, Texas 78712 \\ \and
Friedrich W. Hehl\\
Institute for Theoretical Physics,
University of Cologne\\D-50923 K{\"o}ln, Germany}

\maketitle

\medskip

\begin{abstract}
  Examples in which spacetime might become non-Riemannian appear above
  Planck energies in string theory or, in the very early universe, in
  the inflationary model. The simplest such geometry is metric-affine
  geometry, in which {\it nonmetricity} appears as a field strength,
  side by side with curvature and torsion. In matter, the shear and
  dilation currents couple to nonmetricity, and they are its sources.
  After reviewing the equations of motion and the Noether identities,
  we study two recent vacuum solutions of the metric-affine gauge
  theory of gravity.  We then use the values of the nonmetricity in
  these solutions to study the motion of the appropriate test-matter.
  As a Regge-trajectory like hadronic excitation band, the test matter
  is endowed with shear degrees of freedom and described by a world
  spinor.
\end{abstract}
\bigskip\bigskip

\section{The Case for Metric-Affine Gravity}

Even though Einstein's treatment of spacetime as a Riemannian manifold
appears fully corroborated experimentally, there are several reasons
to believe that the validity of such a description is limited to
macroscopic structures and to the present cosmological era.
Indications \cite{quantum} from the only available finite perturbative
treatment of quantum gravity -- namely the theory of the quantum
superstring -- point to non-Riemannian features on the scale of the
Planck length.  On the other hand, recent advances in cosmogony, i.e.
in the study of the early universe, as represented by the inflationary
model, involve, in addition to the metric tensor, at the very least a
scalar dilaton \cite{inflation} induced by a Weyl geometry, i.e.\ 
again an essential departure from Riemannian metricity.

Allowing minimal departures from Riemannian geometry (i.e.\ from a
$V_{4}$ manifold) would consist in allowing {\it torsion} (i.e.\ a
$U_{4}$) and nonmetricity (i.e.\ an $(L_{4},g)$). {\it Andrzej
  Trautman}, to whom this article is dedicated on the occasion of his
64th brithday, has made important contributions to the study of the
first suggestion \cite{trautman,trautman1}, namely the possibility of
a spacetime with torsion $T^{\alpha}\neq 0$. In this work, we would
like to sketch some of the features relating to the second
possibility, namely to the assumption that {\it spacetime is endowed
  with nonmetricity}\footnote{This is a `positive' paraphrasing of the
  more conventional `negative' assertion, namely that spacetime does
  not fulfill the Riemannian metricity constraint
  $Q_{\alpha\beta}=-Dg_{\alpha\beta}=0$ \cite{nonm}, a wording
  influenced by our Einsteinian conditioning.} $Q_{\alpha\beta}\neq
0$.

We have recently reviewed \cite{PRs} the class of gravitational
theories with such geometries, the {\it Metric-Affine Gauge Theories
  of Gravity} (or `metric-affine gravity' MAG for short). As in any
gauge theory, the geometrical fields of gravity are induced by matter
currents. In Einsteinian gravity, it is the symmetric (Hilbert)
energy-momentum current which acts as a source for the metric field
and the Riemannian curvature.  In MAG, we have, in addition, the {\it
  spin} current and the {\it dilation} plus {\it shear} currents
inducing the torsion and nonmetricity fields, respectively. Both spin
and dilation plus shear are components of the {\it hypermomentum}
current, symmetric for dilation plus shear and antisymmetric for spin.

And yet, there is a rather profound difference between these two
physical features. Special Relativity (SR) is synonymous with
Poincar{\'e} invariance, which includes the Noether conservation of
angular momentum. It would then be relatively straightforward, at the
level of Relativistic Quantum Field Theory (RQFT), to constrain the
kinematics so as to do away with the {\it orbital} part of angular
momentum and thus obtain {\it a conserved spin current}. However, even
this is unnecessary, since through its Pauli-Luba{\'n}ski realization,
{\it spin itself is related to the density of a Poincar{\'e} group
  invariant} and is thus an `absolute' of SR. The conservation of
shear, on the other hand, is not a characteristic of SR and would
require the homogeneous Lorentz group -- or its double-covering group
$SL(2,C)=\overline{SO}(3,1)$ -- to be embedded in the larger
$\overline{SL}(4,R)$. Such switching from Poincar{\'e} to affine
$\overline{A}(4,R)=R^{4}{\;{\rlap{$\supset$}\times}\;}
\overline{SL}(4,R)$ is, however, implied in our having given up the
Riemannian metricity condition, since we have thereby also lost the
presence of the pseudo-orthogonal group as the local symmetry of the
tangent manifold, i.e.\ the local Lorentz frames and with them the
Equivalence Principle, with the direct transition to SR. Indeed, this
is the `meaning' of our basic non-Riemannian ansatz, namely that we
are studying phenomena and situations in which there is no
conventional `flat' SR limit -- either in the small, when approaching
Planck length, or in the early universe, during inflation, within
Planck times from the `seeding' vacuum fluctuation `event'.
Presumably, it is then through a spontaneous breakdown of the local
$\overline{A}(4,R)$ symmetry below Planck energies, down to
Poincar{\'e} invariance, that SR and the Riemannian metricity
condition set in (see \cite{nsbreak,tresbreak} for such examples).
Alternatively, we might be dealing with situations in which {\it the
  dynamics} have led to boundary conditions generating shear currents
-- quadrupolar pulsations of nuclear or hadron matter in the small
\cite{dothan}, e.g., or the Obukhov-Tresguerres {\it hyperfluid}
\cite{hyperfluid} in macroscopic configurations.

\section{World Spinors as Matter Fields}

The unavailability of local Lorentz frames poses no problem in the
context of boson fields. The latter are conventionally represented by
{\it tensors}, i.e.\ linear field representations of $SL(4,R)$. These
become {\it world tensors} in the transition from Special to General
Relativity, i.e.\ nonlinear realisations of the group of local
diffeomorphisms $Di\!f\!f(4,R)$, carried linearly through the
$SL(4,R)_{H}$ holonomic linear subgroup. In RQFT, the fact that tensor
fields are built to carry the action of a group larger than that
allowed by SR, is taken care of through subsidiary conditions, etc.
Thus, the (symmetric) metric tensor density's 10 components
$\sigma^{ij}$, as defined in GR through the action of $SL(4,R)$, e.g.\ 
through
\begin{equation}
  \sigma^{ij}:=2(-g)^{-1/2}\,\delta L/\delta
  {g_{ij}}\,,\label{symmetricem}
\end{equation}
are a good example of a 10-dimensional SL(4,R)-irreducible multiplet
then reducing under SO(1,3) (or SL(2,R)) into $9+1$ -- the
seggregation of the `1' being assured through the removal of the
trace, indeed a Lorentz scalar. In any case, boson fields are
naturally constructed so as to be capable of carrying the action of
$SL(4,R)$, instead of the Lorentz group, whether in a local frame or
holonomically.

The (symmetric) metric tensor density's 10 components are a good
example of a 10-dimensional SL(4,R)-irreducible multiplet then
reducing under SO(1,3) (or SL(2,R)) into 9+1.

This is not true of the conventional fermion fields we use to
represent {\it matter}. These are spin $\vert J\vert=1/2$ field
representations of the double-covering of the Lorentz or Poincar{\'e}
groups, i.e.\ of $Spin(1,3)=SL(2,C)$ or of
$R^{4}{\;{\rlap{$\supset$}\times}\;} SL(2,C)$ and can only carry -- at
best -- nonlinear realisations of $A(4,R)_{H}\subset Diff(4,R)$.  {\it
  Linear action can nevertheless be realised, through the use of
  infinite-component manifields, linear field representations of the
  double-covering of the linear, affine and diffeomorphism groups}
\cite{worldspinors}. Such fields can be used in a Riemannian context
and even in SR, as well as in our present metric-affine geometry. In
the former, they are particularly suited for the description of
hadrons and nuclei, composite objects displaying excitation bands
\cite{dothan}. These phenomenological features have no other
description in the framework of an effective field theory.  Fermionic
hadrons or nuclei are then assigned to spinor manifields -- world
spinors in GR -- and boson excitation bands to the related boson
manifields (`infinitensors'). As to the non-Riemannian scenarios of
super-Planckian energies or of the early universe, spinor manifields
enter naturally in the context of quantum superstring theory.

As field representations of $\overline{SL}(4,R)$, world 
spinors can also be assigned to a local 
$\overline{SL}(4,R)_{A}$ anholonomic frame, as well as 
serving holonomically and carrying the action of 
$\overline{Di\!f\!f}(4,R)\supset \overline{SL}(4,R)_{H}$.
The different spin levels in a world spinor are related 
by the 
\begin{equation}
\vert\delta  J\vert =2
\label{spin2}\end{equation} 
spin-raising and spin-lowering action of the gravitational field. In
its absence, i.e.\ in SR, world spinor manifields reduce to an
(infinite) direct sum of Lorentz spinor fields -- a reduction similar
in principle to that what happens to the $\sigma^{ij}$ tensor in our
example above; moreover, the anholonomic spinor manifields can be
assigned to the more elegant multiplicity-free representations. World
spinor manifields, however, cannot stay in such representations;
transvection of an anholonomic spinor manifield into a world spinor,
using countable-infinite {\it vielbeins}, destroys the
multiplicity-free feature (see \cite{cant}, also Chapter 4 and
Appendices C1-C6 in \cite{PRs}), as exemplified by the Mickelsson
equation \cite{mick}.

\section{Geometrical Fields, Currents and Equations of Motion}

We denote the frame field by 
\begin{equation}
e_{\alpha}=e^{i}{}_{\alpha}\,\partial _{i}
\label{frame}\end{equation}
and the coframe field by 
\begin{equation}
\vartheta^{\beta}=e_{j}{}^{\beta}dx^{j}
\,.\label{coframe}
\end{equation}
The $GL(4,R)$-covariant derivative for a tensor valued $p$-form is
\begin{equation}
D=d+\Gamma_{\alpha}{}^{\beta}\,\rho(L^\alpha{}_{\beta})\wedge\,,
\label{covariantder}\end{equation}
where $\rho$ is the representation of $GL(4,R)$ and
$L^{\alpha}{}_{\beta}$ are the generators; the connection one-form is
$\Gamma_{\alpha}{}^{\beta}=\Gamma_{i\alpha}{}^{\beta}dx^{i}$.  The
nonmetricity is a one-form
\begin{equation}
Q_{\alpha\beta}:=-Dg_{\alpha\beta}\,,\label{nonmetricity}
\end{equation}
the torsion and curvature are two-forms
\begin{equation}
T^{\alpha}:=D\vartheta^{\alpha},
\label{tor}\end{equation} 
\begin{equation}
  R_{\alpha}{}^{\beta}:= d\Gamma_{\alpha}{}^{\beta} -
  \Gamma_{\alpha}{}^{\gamma} \wedge \Gamma_{\gamma}{}^{\beta}\,.
\label{curv}\end{equation}
The Weyl one-form 
\begin{equation}
  Q:=(1/4)\,Q_{\gamma}{}^{\gamma},
\label{weylone}\end{equation}
when subtracted from the nonmetricity, yields the 
{\it traceless} nonmetricity
\begin{equation}
{\nearrow\!\!\!\!\!\!\!Q}_{\alpha\beta}:=Q_{\alpha\beta}-Q\,g_{\alpha\beta}\,.
\label{tracelessq}\end{equation}
The Bianchi identities are
\begin{eqnarray}\label{bianchi}
  DQ_{\alpha\beta}&\equiv& 2R_{(\alpha\beta)}\,,\\ DT^{\alpha}&\equiv&
  R_{\gamma}{}^{\alpha}\wedge\vartheta^{\gamma}\,,\\
  DR_{\alpha}{}^{\beta}&\equiv& 0\,.
\end{eqnarray}
Physics-wise, $Q_{\alpha\beta}$, $T^{\alpha}$ and
$R_{\alpha}{}^{\beta}$ play the role of field strengths.

We now turn to the {\it source-currents} for the fields above. These
will depend on the Lagrangian ($\Psi$ is a matter manifield),
\begin{equation}
  L_{\rm tot}=L_{\rm tot}(g_{\alpha\beta},dg_{\alpha\beta},
  \vartheta^{\alpha}, d\vartheta^{\alpha},\Gamma_{\alpha}{}^{\beta},
  d\Gamma_{\alpha}{}^{\beta},\Psi,D\Psi)\,,\label{lagr1}
\end{equation}which can be
rewritten in a covariantized form as 
\begin{equation}
  L_{\rm tot}= L_{\rm tot}(g_{\alpha\beta},Q_{\alpha\beta},
  \vartheta^{\alpha}, T^{\alpha},R_{\alpha}{}^{\beta},\Psi,D\Psi)\,.
\end{equation} 
Separating the Lagrangian $L_{\hbox{tot}}=V_{\rm MAG}+L$ into
geometrical $V_{\rm MAG}$ and matter $L$ parts, the matter current
three-forms are then given by the Euler-Lagrange functional
derivatives (denoted by $\delta $) of the material piece $L$. We have 
the canonical energy-momentum current
\begin{equation}
  \Sigma_{\alpha}:=\delta L/\delta \vartheta^{\alpha}= \partial L /
  \partial \vartheta^{\alpha} + D(\partial L / \partial T^{\alpha})\,,
\label{canem}
\end{equation}
the hypermomentum current 
\begin{eqnarray}
  \Delta ^{\alpha}{}_{\beta}&:=& \delta L/\delta
  \Gamma_{\alpha}{}^{\beta}= (L^{\alpha}{}_{\beta}
  \Psi)\wedge(\partial L/\partial (D\Psi))\nonumber\\ &+&
  2g_{\beta\gamma}(\partial L/\partial Q_{\alpha\gamma}) +
  \vartheta^{\alpha}\wedge(\partial L/\partial T^{\beta}) + D(\partial
  L/\partial R_{\alpha}{}^{\beta})\,,
\label{canhyp}\end{eqnarray}
and also a related `current', which is a four-form, the (symmetric)
metric energy-momentum, which we used in (\ref{symmetricem}) as an
example of a tensor which reduces under SR, namely
\begin{equation}
  \sigma^{\alpha\beta}:=2 \delta L/\delta g_{\alpha\beta}= 2\partial
  L/\partial g_{\alpha\beta}+ 2D(\partial L/\partial Q_{\alpha\beta})\,.
\label{hilbertem}\end{equation}

Then the field equations turn out to be \cite{PRs}
\begin{eqnarray} 
  \delta L/\delta \Psi &=&0\qquad\qquad (matter)\,,\label{matmat}\\ 
  DM^{\alpha\beta}-m^{\alpha\beta}&=&\sigma^{\alpha\beta}\qquad\quad(0th)\,,\\ 
  DH_{\alpha}- E_{\alpha}&=& \Sigma_{\alpha}\qquad\quad\;(1st)\,,\\ 
  DH^{\alpha}{}_{\beta}- E^{\alpha}{}_{\beta}&= & \Delta
  ^{\alpha}{}_{\beta}\qquad\;\;\,(2nd)\,,
\end{eqnarray}
where we have used the canonical momenta (`excitations'), 
\begin{equation}
M^{\alpha\beta}:=- 2\partial V_{\rm MAG}/\partial Q_{\alpha\beta}, 
\label{h0}\end{equation}
a (three-form) momentum conjugate to the metric field, 
\begin{equation}
H_{\alpha}:=- \partial V_{\rm MAG}/\partial T^{\alpha}\,,
\label{h1}\end{equation}
a (two-form) momentum conjugate to the coframe field and 
\begin{equation}
H^{\alpha}{}_{\beta}:=- \partial V_{\rm MAG}/\partial R_{\alpha}{}^{\beta}\,,
\label{h2}\end{equation}
the (two-form) momentum conjugate to the $GL(4,R)$-connection.

The currents $m^{\alpha\beta}, E_{\alpha}, E^{\alpha}{}_{\beta}$ are
respectively components of the metric energy-momentum, of the
canonical energy-momentum and of the hypermomentum currents,
contributed by the gravitational fields themselves, in $V_{\rm MAG}$ --
the so-called vacuum contributions.

Diffeomorphisms and $GL(4,R)$ invariance yield two Noether identities
\cite{PRs} which, given in their `weak' form, i.e.\ after the
application of the matter equation of motion (\ref{matmat}), become
\begin{equation}
D\Sigma_{\alpha}=
(e_{\alpha}\rfloor T^{\beta})\wedge\Sigma_{\beta} +
(e_{\alpha}\rfloor R_{\beta}{}^{\gamma}) \wedge 
\Delta ^{\beta}{}_{\gamma} -
(1/2)(e_{\alpha}\rfloor Q_{\beta\gamma})\,\sigma^{\beta\gamma}\,, 
\label{noether1}\end{equation}
\begin{equation}
D\Delta ^{\alpha}{}_{\beta} +\vartheta^{\alpha}\wedge\Sigma_{\beta} -
g_{\beta\gamma}\, \sigma^{\alpha\gamma} = 0\,.
\label{noether2}\end{equation}

\section{The OVETH Spherically-Symmetric Vacuum Solution} 

The search for exact solutions to the field equations of MAG is still
in its infancy. First, Tresguerres \cite{tresdil} and, subsequently,
Tucker and Wang \cite{twdil} treated simplified situations, in which
only gravitational {\it dilation} currents represented a departure
from Riemannian geometry.  Recently, a vacuum solution (i.e.\ with
$L=0$) has been found by Obukhov et al.\ \cite{shear1} (`OVETH'), in
which the selection of $V_{\rm MAG}$, however, is such as to provide
for (gravitational) sources of {\it shear}, dilation and spin.  Most
recently, a static vacuum solution with axial symmetry was added to
the set \cite{shear2} (`VTOH').

In terms of irreducible components, in a metric-affine spacetime, 
the curvature has 11 pieces, the torsion 3 and the 
nonmetricity 4 (see App.B in \cite{PRs}). A general quadratic 
Lagrangian (signature $-+++$) can thus be written as
\begin{eqnarray} 
\label{VMAG} V_{\rm MAG}&=&
\frac{1}{2\kappa}\,\left[-a_0\,R^{\alpha\beta}\wedge\eta_{\alpha\beta}
-2\lambda\,\eta+ T^\alpha\wedge{}^*\!\left(\sum_{I=1}^{3}a_{I}\,^{(I)}
T_\alpha\right)\right.\nonumber\\
&+&\left.  2\left(\sum_{I=2}^{4}c_{I}\,^{(I)}Q_{\alpha\beta}\right)
\wedge\vartheta^\alpha\wedge{}^*\!\, T^\beta + Q_{\alpha\beta}
\wedge{}^*\!\left(\sum_{I=1}^{4}b_{I}\,^{(I)}Q^{\alpha\beta}\right)\right]
\nonumber\\&- &\frac{1}{2}\,R^{\alpha\beta} \wedge{}^*\!
\left(\sum_{I=1}^{6}w_{I}\,^{(I)}W_{\alpha\beta} +
  \sum_{I=1}^{5}{z}_{I}\,^{(I)}Z_{\alpha\beta}\right)\,.  
\end{eqnarray} 
Here $\kappa:=2\pi{\ell_{\rm Planck}}^{2}/(hc)$ is the gravitational
and $\lambda$ the cosmological constant, $\eta$ is the volume
four-form, $\eta_{\alpha\beta}:=
{}^{\ast}(\vartheta_{\alpha}\wedge\vartheta_{\beta})$ and
$a_{0-3},$ $b_{1-4},$ $c_{2-4},\,w_{1-6},\,z_{1-5}$ are
dimensionless coupling constants. The antisymmetric and symmetric
components of the curvature are denoted by
$W_{\alpha\beta}:=R_{[\alpha\beta]}$ and
$Z_{\alpha\beta}:=R_{(\alpha\beta)}$, respectively. 

The OVETH solution belongs to a somewhat simplified Lagrangian with 
\begin{equation}
  w_{I}=0\,,\qquad z_{1}=z_{2}=z_{3}=z_{5}=0\,,
\label{constraint}\end{equation}
i.e.\ preserving in $V_{\rm MAG}$ only one component
${}^{(4)}Z_{\alpha\beta}:=R_{\gamma}{}^{\gamma}g_{\alpha\beta}/4$ from
the symmetric part of the curvature, namely the trace, Weyl's
segmental curvature.  In addition, the following constants won't occur 
in the solution and can be put to zero:
\begin{equation}
a_{1}=a_{3}=b_{1}=b_{2}=c_{2}=0\,.
\label{constraint1}\end{equation}
This leaves in $V_{\rm MAG}$ terms involving two pieces of the 
nonmetricity, a {\it shear} 
\begin{equation}
{}^{(3)}Q_{\alpha\beta}=
(4/9)\left(\vartheta_{(\alpha}e_{\beta)}\rfloor \Lambda - 
(1/4)g_{\alpha\beta}\Lambda\right)\,,
\label{qpiece}\end{equation}
with 
\begin{equation}
  \Lambda:=\vartheta^{\alpha}\left(e^{\beta}\rfloor
    {\nearrow\!\!\!\!\!\!\!Q}_{\alpha\beta}\right)\,,
\label{lambda}\end{equation}
and the {\it dilation} 
\begin{equation}
{}^{(4)}Q_{\alpha\beta}=Q_{\gamma}{}^{\gamma}g_{\alpha\beta}/4=
Q\,g_{\alpha\beta}\,.
\label{dil}\end{equation}
Torsion appearing in $V_{\rm MAG}$ is restricted to its vector piece,
\begin{equation}
{}^{(2)}T^{\alpha}=(1/3)\,\vartheta^{\alpha}\wedge T\,,
\label{toison}\end{equation}
with  
\begin{equation}
  T:= e_{\beta}\rfloor T^{\beta}\,.
\label{toisiontrace}\end{equation}

Taking polar Boyer-Lindquist coordinates
$(t,r,\theta,\phi)$ $\equiv$ $(\hat{0},\hat{1},\hat{2},\hat{3})$ and a
Schwarzschild type (static, spherically symmetric and
Minkowski-ortho\-normal) coframe, with one unknown function $f(r)$,
\begin{equation}
  \vartheta ^{\hat{0}} =\, f\, d\,t \,,\quad\vartheta ^{\hat{1}} =\,
  (1/f)\, d\, r\,, \quad\vartheta ^{\hat{2}} =\, r\,
  d\,\theta\,,\quad\vartheta ^{\hat{3}} =\, r\, \sin\theta \, d\,\phi
  \,,\label{coframe1}
\end{equation}
i.e.\ a metric                                               
\begin{equation} 
  ds^2= -f^2\,dt^2+{dr^2}/{f^2} +r^2\left(d\theta^2+\sin^2\theta
    \,d\phi^2\right)\label{schwarz}
  =o_{\alpha\beta}\,\vartheta^\alpha\otimes\vartheta^\beta \,,
\end{equation}
where we have also used the local Minkowski metric
$o_{\alpha\beta}=\hbox{diag}(-1,1,1,1)$, then the three one-forms
$Q,\Lambda,T$ should have the structure
\begin{equation}
  Q=u(r)\,f\,dt\,,\quad\quad\Lambda=v(r)\,f\,dt\,, \quad\quad
  T=\tau(r)\,f\,dt\,.\label{genEug}
\end{equation}

The exact solution is given by the functions
\begin{equation}
  f=\sqrt{1-(2\kappa M/r)+ (\lambda r^{2}/3a_{0}) +
    z_{4}\left[\kappa(k_{0}N)^{2}/2a_{0}r^{2}\right]}\,,
\label{fshear1}\end{equation}
\begin{equation}
  u={k_{0}{N}/ f r}\,,\qquad v={k_{1}{N}/ f r}\,,\qquad \tau= {k_{2}N/
    f r}\,,\label{coul}
\end{equation}
with $M$ and $N$ arbitrary integration constants and the couplings
$k_{0},k_{1},k_{2}$ given by combinations of the
$a_{0},a_{2},b_{3},c_{3},c_{4}$ dimensionless couplings in the
Lagrangian. In addition, $b_{4}$ is constrained by a condition
relating it to the five other couplings \cite{shear1}.
The nonmetricity is thus given by
\begin{equation}
  Q^{\alpha\beta}=(1/r)\left[k_{0}No^{\alpha\beta} + (4/9)\,k_{1}N
    \left(\vartheta^{(\alpha}e^{\beta)} \rfloor
      -(1/4)\,o^{\alpha\beta}\right) \right] \ dt
\label{non1}\end{equation}
and the torsion by 
\begin{equation}
  T^{\alpha}= \left(k_{2}N/3\,r\right) \vartheta^{\alpha}\wedge
  dt\,.
\label{tor1}
\end{equation}
The integration constant $M$ is the Schwarzschild mass, $k_{0}N$ a
dilation, $k_{1}N$ a (traceless) shear and $k_{2}N$ a spin charge.
For $N=0$ and $a_{0}=1$, one recovers the Schwarz\-schild-deSitter
solution in GR.

For our purposes, we note that the vacuum solution's nonmetricity,
which will couple to a test particle's shear, in this
spherically-symmetric case, represents a $1/r$ potential.

\section{The VTOH Axially-Symmetric Vacuum Solution}

Still using the gravitational Lagrangian (\ref{VMAG}), with the
simplifications (\ref{constraint}) and (\ref{constraint1}), and the
same polar coordinates, VTOH \cite{shear2} posit a Kerr type solution
\begin{eqnarray}
  \vartheta ^{\hat{0}} &=&\,({ A }/{ B })^{\;1/2}\, \left(\,d\,
    t -j_0\sin ^2\theta\,d\phi\,\right),\nonumber\\ \vartheta
  ^{\hat{1}} &=&\,({ A /{ B }})^{-1/2}\;d\,r,\nonumber\\ 
  \vartheta ^{\hat{2}} &=&\,({ B }/
  f)^{\;1/2}\;d\,\theta,\nonumber\\ \vartheta ^{\hat{3}}
  &=&\,({ B }/ f)^{-1/2}\,\sin\theta\, \left[\,- j_0\, d\, t
    +\left(\,r^2+j_0^2\,\right) d\phi\,\right],\label{axframe}
\end{eqnarray}
where $ A = A (r)\,, \, B = B (r,\theta)\,,\, f=f(\theta)$, and
$j_0$ is a constant.  The three residual one-forms $Q$, $T$, $\Lambda$
of Sec.\ 4 are now replaced by expressions involving three functions
$u(r,\theta)\,,\, v(r,\theta)\,,\, \tau(r,\theta)$ appearing in the third
and fourth irreducible components of nonmetricity and in vector
torsion,
\begin{eqnarray}
  Q_{\alpha\beta}&=&\left[u(r,\theta)\;o_{\alpha\beta}
    +({4}/{9})\,v(r,\theta)\,\left(\vartheta_{(\alpha}e_{\beta )}\rfloor
      -(1/4)\,o_{\alpha\beta}\right)\right]\vartheta^{\hat{0}}\,,
\label{nonmet}\\T^\alpha&=&({1}/{3})\,\tau(r,\theta)\;
\vartheta^\alpha\wedge\vartheta^{\hat{0}}\,,\label{tor2}
\end{eqnarray}
with the solutions, 
\begin{eqnarray}
  u&=&k_{0}Nr/( A  B )^{1/2}, \\ v&=&k_{1}Nr/( A  B )^{1/2}, \\ 
  \tau&=&k_{2}Nr/( A  B )^{1/2}\,,
\end{eqnarray}
and 
\begin{eqnarray}
 A  &=&r^2+j_0^2-2\kappa Mr-({{\lambda}/ 3a_0})\,
r^2\left(r^2+j_0^2\right) +z_4\kappa (k_0N)^2/ (2a_0)\,,\label{sol2a}\\ 
 B  &=&\,r^2 +j_0^2\cos^2\theta,\label{sol2b}\\ 
f&=&\,1+({{\lambda}/ 3a_0})\,j_0^2\cos ^2\theta\,.\label{sol2c}
\end{eqnarray}
Here the $k_{0}, k_{1}, k_{2}$ are functions of the couplings $a_{0},
a_{2}, b_{3}, c_{3}, c_{4}$ (the same ones as in Sec.\ 4) and, again,
the same constraint relates $b_{4}$ to $k_{0}, k_{1}, k_{2}, c_4$.
Physically, $M$ and $j_{0}$ represent the Schwarzschild mass and the
Kerr angular momentum. For vanishing $j_0$, we recover the OVETH
solution of Sec.\ 4.

\section{The Test-Particle in the OVETH and VTOH Solutions}

The Noether identities (\ref{noether1},\ref{noether2}) already provide
important information with respect to the behaviour of test-matter in
MAG. Obukhov \cite{obukhov}, generalizing a corresponding result
\cite{Meyer,He5} from Riemann-Cartan spacetime, has recast equation
(\ref{noether1}) in the form (quantities with tilde denote the
Riemannian parts),
\begin{equation}
  \widetilde{D}\left[\Sigma_\alpha+\Delta^{\beta\gamma}
    \left(e_\alpha\rfloor\diamondsuit
      \Gamma_{\beta\gamma}\right)\right]
  +\Delta^{\beta\gamma}\!\wedge\!\left(\pounds_{e_\alpha}\diamondsuit
    \Gamma_{\beta\gamma}\right)=\tau^{\beta\gamma}\!\wedge\!
    \left(e_\alpha\rfloor\widetilde{R}_{\gamma\beta}\right)
    \,,\label{Yuri}
\end{equation}
where $\diamondsuit\Gamma_{\beta\gamma}:=\Gamma_{\beta\gamma}-
\widetilde{\Gamma}_{\beta\gamma}$ denotes the non-Riemannian part of
the connection. The expression on the right hand side of (\ref{Yuri})
represents the Mathisson-Papapetrou force density of GR for matter
with spin $\tau^{\beta\gamma}:=\Delta^{[\beta\gamma]}$. For $\Delta
^{\beta\gamma}=0$, the equation of motion becomes
$\widetilde{D}\Sigma_{\alpha}=0$, i.e.\ without dilation, shear and
spin `charges' the particle follows {\it Riemannian geodesics}, irrespective
of the composition of $V_{\rm MAG}$. Thus, we have to use as test
matter only configurations which carry dilation, shear or spin
charges, whether macroscopic or at the quantum particle level. At the
latter, the hadron Regge trajectories provide adequate test matter, as
world spinors with {\it shear}.

In the world spinor equation, when written anholonomically, the
$GL(4,R)$ Lie-algebra-valued connection
$\Gamma_{\alpha}{}^{\beta}[\rho(L^{\alpha}{}_{\beta})]_{N}{}^{M}$,
acting on the component $\Psi^{N}(x)$, parallels the action of the
same $GL(4,R)$ generators (symmetric
$g_{\gamma(\alpha}L^{\gamma}{}_{\beta)}$ for shear and dilations) in
the expression for the original Noether current of hypermomentum for
the matter Lagrangian, $\Delta ^{\alpha}{}_{\beta}=
[(L^{\alpha}{}_{\beta})_{N}{}^{M} \Psi^{N}]\wedge [\partial L/\partial
(D\Psi)^{M}]+\cdots$, where they enter in writing the variation of the
matter field.  This is just a reflection of the {\it universality} of
gauge couplings, in which gauge fields are coupled to conserved
currents.

An identity (Eq.(3.10.8) in Ref.\ \cite{PRs}) expresses the components
of the connection one-form as a linear combination of the components
of the Christoffel symbol (of the first kind), the object of
anholonomity $C^\alpha:=d\vartheta^\alpha$, the torsion $T^\alpha$ and
the nonmetricity $Q^{\alpha\beta}$,
\begin{equation}\Gamma_{\gamma\alpha\beta} = ({1/2})\,
  [ \partial_{\lbrace \gamma}g_{\beta\alpha \rbrace } + C_{\lbrace
    \gamma\beta\alpha\rbrace } - T_{\lbrace \gamma\beta\alpha\rbrace
    }+ Q_{\lbrace \gamma\beta\alpha\rbrace }] \,,\label{6.2}
\end{equation}
where the $\{\}$ are Schouten braces \cite{schouten}. It is through 
this replacement that we get in the matter equation (\ref{matmat}) 
the action of the nonmetricity field. This can also be rewritten as:
\begin{equation}
  \Gamma_{\alpha\beta}=[V_{4}\hbox{-terms}] + [U_{4}\hbox{-terms}] +
  (1/2)\,Q_{\alpha\beta} +(e_{[\alpha}\rfloor
    Q_{\beta]\gamma})\vartheta^{\gamma}\,.
\label{6.3}\end{equation}

Turning now to the two vacuum solutions with nonmetricity, and leaving
out dilations ($k_{0}=0$), we find, with $G:=(1/9)\,k_{1}N$,
\begin{equation}
  {}^{(3)}{Q}_{\alpha\beta}= 4G\,r\,( A  B )^{-1/2}
\left(\vartheta_{(\alpha}e_{\beta )}\rfloor-(1/4)\,o_{\alpha\beta}\right)
\vartheta^{\hat{0}}\,,
\label{6.4}\end{equation}     
or 
\begin{equation}
  \left({}^{(3)}{Q}\right)_{\alpha\beta}= { G\,r\over
    ( A  B )^{1/2}}
  \left(\begin{array}{rccc}-3\vartheta^{\hat{0}}&2\vartheta^{\hat{1}}&2
      \vartheta^{\hat{2}}&2\vartheta^{\hat{3}}\\2\vartheta^{\hat{1}}
      &-\vartheta^{\hat{0}}&0&0\\ 2 \vartheta^{\hat{2}}&0&
      -\vartheta^{\hat{0}}&0\\2\vartheta^{\hat{3}}
      &0&0&-\vartheta^{\hat{0}}
\end{array}\right) 
\label{6.5}\,.\end{equation}
Particularly simple are the diagonal elements,
\begin{eqnarray}
 {}^{(3)}{Q}_{\hat{0}\hat{0}}/3&=& {}^{(3)}{Q}_{\hat{1}\hat{1}}=
 {}^{(3)}{Q}_{\hat{2}\hat{2}}= {}^{(3)}{Q}_{\hat{3}\hat{3}}\nonumber
\\& =& -G\,r\left(dt-j_{0}\sin^{2}\theta\,d\phi\right)/
\left(r^{2}+j_{0}^{2}\cos^{2}\theta\right)\label{6.6}\,,
\end{eqnarray}
which reduce to the spherically symmetric $-Gdt/r$ for $j_{0}=0$. This
is then the static potential entering, via (\ref{6.3}), the world spinor
equation, a $1/r$ potential with a centrifugal cut-off at small $r$.

Applying this potential to a Dirac or Bargmann-Wigner equation,
written for any component in the `flat' equation (4.5.1) of Ref.\ 
\cite{PRs}, we get in the spherically-symmetric case, a hydrogen-like
relativistic spectrum, thereby superimposed on every state in the
diagonal, with multiplicities growing with the Bargmann-Wigner spin
value.  The resulting world spinor is thus much more populated, but
not yet in the off-diagonal sectors of Fig.3 in Ref.\ \cite{PRs}.

In the axial-symmetric case, $j_{0}\neq0$, and we thus have, 
in addition, the $\vert\delta J\vert=2$ action, reaching into
the {\it off-diagonal} sectors and partially filling them.  

The energy-spectrum will follow. The generic world spinor corresponds
to $\overline{SA}(4,R)$ representations in class IIA (see Appendix C5
of Ref.\ \cite{PRs}), i.e.\ with no kinematical constraint on the mass
spectrum. Let us take, for instance, a (hadronic) linear
$M^{2}=aJ_{0}+b$ as our free world spinor. We shall now have, in the
simplest (lowest state $J=1/2$) case, a superposition of the
(Dirac-relativistic) hydrogen-like gravitational excitations {\it due
  to nonmetricity}, onto the linear spectrum of the flat limit (`free')
  manifield.  Next, we would have to solve the Bargmann-Wigner
  equation for $J=5/2$ in this hydrogen-like potential, etc..

\pagebreak

\centerline{******************}

\end{document}